\documentclass[a4paper]{jpconf}
\usepackage{graphicx}
\begin{document}
\title{M(atrix) model interaction with 11D supergravity}

\author{Igor A. Bandos}

\address{Department of
Theoretical Physics, University of the Basque Country
 UPV/EHU,
P.O. Box 644, 48080 Bilbao, Spain and IKERBASQUE, Basque Foundation for Science, Bilbao, Spain}

\ead{igor\_bandos@ehu.es, bandos@ific.uv.es}

\begin{abstract}
We present the equations of motion for multiple M0--brane (mM0) system in an arbitrary curved supergravity superspace which generalizes the M(atrix) model equations for the case of arbitrary supergravity background. Although these were obtained in the frame of superembedding approach to mM$0$, we do not make a review of this approach in this contribution but concentrate discussion on  the structure of the equations.

\end{abstract}

\section{Introduction}

M(atrix) model conjecture \cite{Banks:1996vh} attracted - and still attracts-  much attention as a relatively simple approach to a nonperturbative description of String/M-theory, the most promising candidate on the r\^ole of quantum theory of gravity and of the unified theory of all fundamental interactions. The basic Lagrangian used to study that 'M(atrix) theory' was the one obtained by dimensional reduction of $D$=$10$ $U(N)$ supersymmetric Yang--Mills theory (SYM) down to $d$=$1$. This is treated as a (very low energy and gauge fixing) description of the dynamical system of $N$  D$0$--branes (D-particles or Dirichlet particles) moving in flat target $D=10$ type IIA superspace.

D0-brane  is the simplest representative of the family of D$p$-branes (Dirichlet $p$--branes with $p$=$1$ for string, $p$=$2$ for membrane etc.) defined as hypersurfaces where string can have its ends \cite{Sagnotti,Horava}. As String theory is a theory of gravity, a `frozen' hypersurface cannot exist in its frame so that  D$p$--branes are dynamical supersymmetric extended objects. Their  equations of motion can be obtained from actions defined on their ($p$+$1$) dimensional worldvolumes $W^{p+1}$ embedded in target $D$=$10$ type II superspace $\Sigma^{(10|32)}$ with ten bosonic and $32$=2$\times$16 fermionic dimensions.

In general, the interaction between  constituents of a multiple D$p$ (mD$p$) system is due to the strings stretched between different D$p$--branes. At low energy the massive string excitations can be ignored and the string is described by its massless modes which are the fields of SYM multiplet. When a set of N nearly coincident Dp-branes is considered, the strings scratched between different branes give rise to the fields of $U(N)$ SYM multiplet \cite{Witten:1995im}. This explains why the degrees of freedom of the multiple Dp-brane system are the ones of the maximally supersymmetric (16 supersymmetries) $U(N)$ SYM multiplet. However, being the objects of the type II String Theory, multiple Dp-brane system should actually possess 32 component supersymmetry, the 16 of which are realized linearly (and preserved by the ground state of mD$p$ system and in the SYM description) while the other 16 supersymmetries are realized nonlinearly (spontanousely broken by the mDp system). The mechanism of enhancement of 10D ${\cal N}=1$ supersymmetry (with 16 generators) till 10D type II supersymmetry (with 32 generators) by passing from the SYM to nonlinear action containing Dirac--Born--Infeld and  Wess--Zumino terms  is well understood for the case of a single Dp-brane (see \cite{B+T=Dpac} and \cite{Dima99,IB09:M-D} for more references).

Thus the $d=1$ ${\cal N}=16$ $U(N)$ SYM action provides a very-low-energy description of the nearly coincident mD$0$ system in flat target type II superspace and just this Lagrangian was used to describe the M(atrix) theory of \cite{Banks:1996vh}. The restoration of the 11D Lorentz invariance starting from $1d$ ${\cal N}=16$  SYM Lagrangian was also considered in \cite{Banks:1996vh} and allowed to treat this as the Matrix model Lagrangian in flat 11D supergravity background. However, till very recent time, the matrix model equations were known for  very few particular non-flat supergravity backgrounds including pp-waves \cite{BMN} and the light--like linear dilaton background \cite{Verlinde+C+S=2005}.

The matrix model equations for an arbitrary supergravity background, which we will discuss here, have been obtained in \cite{mM0=PRL} by developing superembedding approach to the system of multiple M$0$-branes.
Having said that, we have to explain briefly what is M$0$-brane and what is  the superembedding approach to super-$p$-branes (although we will not review this in the present contribution but just present the results obtained in its frame).

Single M$0$--brane is just the eleven dimensional  massless superparticle \cite{B+T=Dpac}.  It is the simplest representative of the family of 11D M-branes also including supermembrane (M2-brane) \cite{BST1987,bpstv} and M-theory 5-brane (M5-brane) \cite{hs2,blnpst,schw5}. M$0$-brane can also be called M-wave; the name multiple gravitons was used  in \cite{YLozano+=0207} for the (bosonic limit of the) multiple M$0$ (mM$0$) system.

The superembedding approach to supersymmetric extended objects, (super-)p-branes
\cite{bpstv,hs96,hs2,Dima99,IB09:M-D}, following the so--called STV (Sorokin--Tkach--Volkov) approach to superparticles and superstrings \cite{stv} (see \cite{Dima99} for review and more refs)  describes $p$-branes in terms of embedding of its {\it worldvolume
superspace} ${\cal W}$   into the {\it target superspace} $\Sigma$. For the case of M$0$-brane these are the $d$=$1$ ${\cal N}$=$16$ superspace ${\cal W}^{(1|16)}$ and the $D$=$11$ ${\cal N}$=$1$ superspace $\Sigma^{(11|32)}$, respectively. The embedding of ${\cal W}$ into $\Sigma$ obeys the so-called {\it superembedding equation} which, in the case of higher dimensional $p$-branes (with also sufficiently high co-dimension $D$-$p$) specifies  completely the $p$--brane dynamics, {\it i.e.} contains $p$--brane  equations of motion among its consequences. This on-shell nature of the superembedding equation, which holds, in particular, for all the above mentioned  M-branes, makes possible to use the superembedding approach to derive equations of motion for new branes (as it was with M5-brane \cite{hs2}) and also for multi--brane systems (although this presently is developed for multi-particles and multi-wave cases only \cite{IB09:D0,mM0=PLB,mM0=PRL})\footnote{The boundary fermion approach  \cite{Howe+Linstrom+Linus}, describing mDp systems on a 'minus one-quantization level', also uses a version of superembedding approach, but with nonstandard superspace including boundary fermion directions.}.

\section{Basic variables and Abelian ($U(1)$) part of the Matrix model equations}

As we have already said, the basic action staying beyond the M(atrix) theory of \cite{Banks:1996vh} is  the dimensional reduction of
the maximal, $D$=$10$ ${\cal N}$=$1$, $U(N)$ SYM theory down to $d=1$. In such a dimensional reduction the
spacial components of the $u(N)$ valued SYM potential give rise to nanoplet of anti-hermitian $N\times N$ matrices of scalar fields $\hat{\bf X}{}^i$, 
while the D=10 Majorana--Weyl fermionic field of the 10D SYM gives rise to 16 $u(N)$ valued fermionic matrices  $\hat{\bf \Psi}_q$,
$q=1,\ldots, 16$.

One can extract the trace part of the above matrix fields thus separating the variables describing the center of mass (center of energy) motion and the relative motion of the constituents of the multiple D$0$-brane (mD$0$) system,
\begin{eqnarray}
\label{hX=hx+X} \hat{\bf X}^i&=&\hat{x}^i(\tau) \, I_{_{N\times N}}+ {\bf X}^i(\tau) \; , \qquad
 i=1,...,9\; , \qquad  tr ({\bf X}^i)=0\; , \qquad \\ \label{hP=hp+P}
\hat{\bf \Psi}_q &=& \hat{\psi}_q (\tau)  I_{_{N\times N}} + {\bf \Psi}_q(\tau)\; , \qquad  q=1,...,16\, , \quad\;\; tr({\bf \Psi}_q)=0\; . \qquad
\end{eqnarray}
The Abelian center of energy variables $\hat{x}^i(\tau)$, $\hat{\psi}_q (\tau)$ are the same as describing a single D$0$ after fixing the so--called static gauge. The same variables are used also for the gauge fixed description of M$0$-brane, while before the gauge fixing, the Lorentz (and/or diffeomorphism) covariant description of a single M$0$ brane is given in terms of the $\Sigma^{(11|32)}$ coordinate functions
\begin{eqnarray}
\label{hZ=hx,hth}
\hat{Z}^M(\tau) = (\hat{x}^\mu(\tau)\, , \, \hat{\theta}^{\check{\alpha}}(\tau))\; ,\qquad \mu=0,1,\ldots, 9,10\; , \qquad \check{\alpha} = 1,\ldots, 32
\; . \qquad
\end{eqnarray}
The number of bosonic coordinate functions in the covariant description of D$0$-brane is 10 ({\it versus}  11 for M$0$, Eq. (\ref{hZ=hx,hth})). The similarity  of the variable used for the gauge fixed description is due to that the D$0$--brane, which is the massive 10D ${\cal N}=2$ superparticle (see \cite{dA+L82} for the first appearance of its D=6 counterpart) can be obtained by dimensional reduction of the 11D massless superparticle in $D=11$, which is to say of the M$0$-brane \cite{B+T=Dpac}.

However, the gauge fixed description of M$0$-brain has a higher residual symmetry: the SO($9$) residual bosonic symmetry of the gauge fixed description of D$0$--brane is enlarged in this case  up to SO(1,1)$\otimes$SO(9)
group\footnote{Actually, the residual symmetry of the gauge fixed M$0$-brane is $[SO(1,1)\otimes SO(9)]\subset\!\!\!\!\!\!\times K_9$, see \cite{IB07:M0,IB09:M-D} and refs therein, but the $K_9$ part is inessential for our discussion here.} so that the basic fields and parameters are also characterized by their SO(1,1) weight. In particular, a parameter of the $\kappa$--symmetry \cite{dA+L82,W.S.83} (which can be identified with worldline supersymmetry \cite{stv} and reflects a part of spacetime supersymmetry which is preserved by M$0$-brane ground state) has SO(1,1) weight $1$ which we mark by sign superscript in its notation: $\epsilon^{+q}$ \cite{mM0=PLB,mM0=PRL}. Clearly the SO(1,10) spinorial parameter $\varepsilon^\alpha (x)$ (and $\varepsilon^\alpha (Z)$) of the 11D  spacetime (superspace) supersymmetry is inert under the above SO(1,1), so that when describing the spacetime supersymmetry preserved by M$0$-brane ({\it i.e.} linearly realized in the M$0$--brane model), $\hat{\varepsilon}{}^\alpha:= \varepsilon^\alpha (\hat{Z}(\tau))$, it is expressed through the above $\kappa$--symmetry parameter $\epsilon^{+q}$ by using the $16\times 32$ matrix  $v^{-\alpha}_q$  which carries  the $SO(1,1)$ weight $-1$,
\begin{eqnarray}\label{susy=1/2susy}
& \hat{\varepsilon}{}^{\alpha}= \epsilon^{+q} v^{-\alpha}_q
 \;  , \qquad \alpha =1,\ldots, 32\; , \qquad q =1,\ldots, 16 \; .
\end{eqnarray}
This rectangular {\it spinor moving frame} matrix carrying the  $Spin(1,10)$ index $\alpha=1,\ldots, 32$
and $Spin(9)$ index $q=1,\ldots, 16$
is constrained to obey
\begin{eqnarray}\label{M0:v-v-=u--}
 v_{q}^- {\Gamma}_{ {a}} v{}_{p}^- = \; u_{ {a}}{}^{=} \delta_{qp}\; , \qquad
 & 2 v_{\alpha}{}_{q}^{-}
 v_{ \beta}{}_{q}^{-}{}= {\Gamma}^{a}_{ {\alpha} {\beta}} u_{ {a}}{}^{=}\; , \qquad
\end{eqnarray}
where ${\Gamma}^{a}_{ {\alpha} {\beta}} := {\Gamma}^{a}_{ {\alpha}}{}^{\gamma} C_{ {\gamma} {\beta}}$ are 11D Dirac matrices contracted with 11D charge conjugation matrix. The vector $u_{ {a}}{}^{=}(\tau)$ in (\ref{M0:v-v-=u--}) is light-like (as a consequence of (\ref{M0:v-v-=u--})). Actually, it is proportional to the momentum of the M$0$-brane, the fact which can be expressed by stating that
\begin{eqnarray}\label{M0:E+a=u-a}
 \hat{E}_{\#}^a:= D_{\#}\hat{Z}^M(\tau) E_M^a(\hat{Z})= u_{ {a}}{}^{=}(\tau)\; ,
\end{eqnarray}
where  $D_{\#}$ is covariant derivative on the worldline, $D_{\#}\hat{Z}^M= e_{\#}^\tau\partial_\tau \hat{Z}^M(\tau)$ and $E_M^a(Z)$ are the component of the target superspace supervielbein form $E^a=dZ^ME_M^a(Z)$. (Notice that in our notation the subscript plus index is equivalent to superscript  minus one and {\it vice versa}). The supervielbein form of the curved 11D superspace $E^A=(E^a,{E}^\alpha) =dZ^ME_M^A(Z)$  carries  degrees of freedom of the 11D supergravity supermultiplet if it obeys the superspace supergravity constraints \cite{CremmerFerrara80,BrinkHowe80}. The most important of these are collected in $T^{{a}}:= DE^{{a}} =
-i{E}^\alpha \wedge  {E}^\beta \Gamma^{{a}}_{\alpha\beta}$.

One can consider $u_{ {a}}{}^{=}(\tau)$ as a part of {\it moving frame} attached to the M$0$ worldline,
\begin{eqnarray}\label{u++u++=0}
&\left({u_a^{\#}+u_a^{=}\over 2}\, , \, u_a^{i} \, , \, {u_a^{\#}-u_a^{=}\over 2}\right)\;  \in \; SO(1,D-1)\Leftrightarrow  \;
\cases{ u_{ {a}}^{=} u^{ {a}\; =}=0\; , \quad u_{ {a}}^{=} u^{ {a}\; i}=0\; ,
\quad
  u_{ {a}}^{\;=} u^{ {a}\;  \# }= 2\; , \cr  u_{ {a}}^{\# } u^{ {a}\; \#
}=0\; ,  \quad
 u_{{a}}^{\;\#} u^{ {a} i}=0\, , \quad   u_{ {a}}^{ i} u^{ {a} j}=-\delta^{ij}\; .} \;
\end{eqnarray}
The M$0$ equations of motion can be conveniently written with the use of the space-like  moving frame vector $u_{ {a}}^{ i} $ (in (\ref{u++u++=0})) and the set of spinors $v_{{\alpha}p}{}^-$ ('square roots' of $u_{ {a}}^{=}$ in the sense of (\ref{M0:v-v-=u--})) as
 \begin{eqnarray}
\label{M0:Eqsm}
 D_{\# } \hat{E}_{\# }{}^{  {a}}\, u_{ {a}}{}^i=0 \; , \qquad \hat{E}_{\# }{}^{ {\alpha}}v_{ {\alpha}p}{}^-:= D_{\#}\hat{Z}^M(\tau) E_M{}^{ {\alpha}}(\hat{Z})v_{ {\alpha}p}{}^-=0
\; . \qquad
\end{eqnarray}
These equations of motion also imply that $v_{\alpha q}^{\; -}$ and  $u_a^{=}$ are covariantly
constant,   $Dv_{\alpha q}^{\; -} =0$ and  $Du_a^{=}=0$ (see \cite{mM0=PLB,mM0=PRL} for more detail).

The moving frame variables (\ref{u++u++=0}) can be also used to extract, in the covariant manner, different projections of the supergravity fluxes to the worldline. The following projections of the 4-form field strength ($F_4=dC_3$), gravitino field strength $T_{ab}{}^\alpha= T_{[ab]}{}^\alpha$ and Riemann tensor of supergravity $R_{dc\;
ba}=R_{[dc]\;
[ba]}$
\begin{eqnarray} \label{M0:Fluxes}\label{M0:Fluxes}
 & \hat{F}_{\# ijk}:= F^{{a}{b}{c}{d}} (\hat{Z}) u_{{a}}
{}^{=}u_{{b}}{}^{i}u_{{c}}{}^{j}u_{{d}}{}^{k}\; , \qquad   \hat{R}_{\# ij \#}:= R_{dc\;
ba}(\hat{Z})u^{d=}u^{ci} u^{bj}u^{a=} \; ,  \qquad \nonumber
\\ \label{M0:fFluxe} & \hat{T}_{\#\, i\, +q}
:=T_{{a}{b}}{}^{\alpha}(\hat{Z})\,  v_{{\alpha}q}^{\; -}\,
u_{{a}}^{=}u_{{b}}^i\;    \qquad
\end{eqnarray}
play a special r\^ole: only they influence the closure of the worldline supersymmetry algebra\footnote{This is tantamount to saying that their superspace generalizations enter the description of the geometry of the worldline superspace ${\cal W}^{(1|16)}$ and normal bundle over it; see \cite{mM0=PRL}.}.

In our superembedding approach \cite{mM0=PLB,mM0=PRL} {\it the center of energy motion of the multiple M$0$ system is described by the same set of equations (\ref{M0:Eqsm}) and characterized by the same supersymmetry properties as a single M$0$}. Then it is natural to expect that the supergravity fluxes can enter the equations of relative motion of the mM$0$ system only through  (\ref{M0:fFluxe}). This is indeed the case.

\section{Non-Abelian ($SU(N)$) part of the Matrix model equations 
in an arbitrary $11D$ supergravity background}

In the light of relation between M$0$ and D$0$ brane, it is natural to expect that the relative motion of the constituents of the mM$0$ system is described by the nanoplet of bosonic traceless matrices ${\bf X}{}^i$ and 16-plet of fermionic traceless matrices ${\bf \Psi}_q$ introduced in (\ref{hX=hx+X}) for the case of D$0$-branes. The superembedding approach of \cite{mM0=PLB,mM0=PRL} imposes on these fields a set of dynamical equations which includes the following fermionic equation
\begin{eqnarray}\label{M0:DtPsi=}
& \dot{\Psi}_{q}=- {1\over 4} \gamma^i_{qp} \left[ {\bf X}^i\, , \,
\Psi_{p} \right]   + {1\over 24}  \hat{F}_{\# ijk} \gamma^{ijk}_{qr}\Psi_{r}
 - {1\over 4}  {\bf X}^i\hat{T}_{\# i \, +q}\, ,
\end{eqnarray}
as well as Gauss constraint
\begin{eqnarray}\label{M0:XiDXi=}
& \left[ {\bf X}^i\, , \, \dot{\bf X}^i\,  \right] = 4i \left\{\Psi_{q}\, , \, \Psi_{q} \right\}  \;  \qquad
\end{eqnarray}
and proper bosonic equation of motion
\begin{eqnarray}\label{M0:DDXi=}
\ddot{\bf X}^i &=
{1\over 16} \left[ {\bf X}^j , \left[ {\bf X}^j ,  {\bf X}^i\,  \right]\right]+ i\gamma^i_{qp} \left\{\Psi_{q}\, , \, \Psi_{p} \right\} +   {1\over 4}  {\bf X}^j \hat{R}_{\# j\; \# i} +
{1\over 8} \hat{F}_{\# ijk} \left[ {\bf X}^j , \, {\bf X}^k\,  \right] -2i \Psi_{q}\hat{T}_{\# i \, +q}  \, . \;
\end{eqnarray}
Here, for the safe of simplicity, we have used  the dot notation for the covariant time derivative,
so that $\;\dot{\Psi}_{q}=D_{\#}\Psi_{q}$, $\;\dot{\bf X}^i:=D_{\#}{\bf X}^i$, $\;\ddot{\bf X}^i:=D_{\#}D_{\#}{\bf X}^i$.

The third term in the {\it r.h.s.} of the bosonic equation (\ref{M0:DDXi=}) shows that, generically, the field ${\bf X}{}^i$ is massless with the mass matrix generated by the special projection of Riemann tensor in  (\ref{M0:fFluxe}). The last, fifth term describes the contribution of the fermionic flux.
The most interesting forth term, $\hat{F}_{\# ijk}[ {\bf X}^j,{\bf X}^k]$, describes the `dielectric coupling'  characteristic for the Emparan-Myers `dielectric brane effect' \cite{Emparan:1997rt,Myers:1999ps}. It is essentially non-Abelian, which is also true for the first and second terms in the {\it r.h.s.} of  (\ref{M0:DDXi=}), although these two are also present in the case of flat background without fluxes and  in 1d dimensional reduction of 10D SYM. 

In this flat 11D superspace case Eqs. (\ref{M0:DtPsi=}), (\ref{M0:XiDXi=}) and (\ref{M0:DDXi=}) also coincide  with the equations describing mD$0$ system in flat 10D type IIA superspace \cite{IB09:D0}\footnote{The dimensional reduction of the curved superspace mM$0$ equations was not studied yet; it might provide a simple way to reproduce the coupling of mD$0$ to type IIA supergravity fluxes.}. However, when discussing mM$0$ system we have to define also the $SO(1,1)$ weight of these matrix valued  fields (the notion which did not appear when mD$0$ system is considered). Superembedding approach \cite{mM0=PLB} suggests to choose the weight of the bosonic and fermionic matrix field to be (-2)  and  (-3), respectively,
\begin{eqnarray} \label{X=X--}
{\bf X}{}^i= {\bf X}{}^{=i}\equiv {\bf X}{}^{--i} \; \qquad {\bf \Psi}_q:= {\bf \Psi}^{=-}_{q}\equiv {\bf \Psi}^{---}_{q}
  \; .    \qquad
\end{eqnarray}
Also the covariant time derivative $D_{\#}=D_{++}\equiv D^{--}$ carries the negative $SO(1,1)$ weith (-2).  As it was noticed in \cite{mM0=PRL}, these observations, considered together with that only the projections of fluxes presented in  (\ref{M0:fFluxe}) may influence the dynamics, makes the structure of the equations for the relative motion of the m$M$0 constituents quite rigid: very few terms are allowed by $SO(1,1)\times SO(9)$ symmetry in addition to the ones which were found as a result of calculations. This restricts possible deformations/nonlinear generalizations of Eqs. (\ref{M0:DtPsi=}), (\ref{M0:XiDXi=}) and (\ref{M0:DDXi=}), at least when the center of mass motion of the mM$0$ system is  not deformed.

\section{Conclusion and discussion}
\setcounter{equation}{0}

In this contribution we have presented the  Matrix model equations in an arbitrary $D$=$11$ supergravity background which were obtained from the superembedding approach to multiple M$0$-brane (mM$0$) system \cite{mM0=PLB,mM0=PRL}. The set of these equations splits naturally on an Abelian ('$U(1)$') part, describing the center of energy motion of the mM$0$ system, and a non-Abelian  ('$SU(N)$') part  containing equations for anti-hermitian traceless $N\times N$ matrices ${\bf X}{}^i$, describing the relative motion of the mM$0$ constituents, and their superpartners ${\bf \Psi}_q$. The equations describing the center of energy motion of the mM$0$ system are the same as equations of motion for a single M$0$-brane, which implies that the center of energy moves on light--like geodesic in a (generically curved) 11D spacetime. The flat target superspace limit of the equations describing the relative motion of the mM$0$ constituents can be also obtained by making dimensional reduction of the $D$=$10$ $SU(N)$ SYM model down to $d$=$1$. However, our equations describe the mM$0$ system in an arbitrary supergravity superspace thus generalizing the Matrix model equations for an arbitrary supergravity background. The right hand sides of these equations contain the contributions from the supergravity fluxes, this is to say, form the antisymmetric tensor field strength $F_{abcd}(x)$, gravitino field strength $T_{ab}{}^\alpha(x)$ and Riemann tensor $R_{ab\; cd}(x)$ of the 11D supergravity. The natural application  is to use our general equations to obtain the Matrix model in physically interesting backgrounds such as  $AdS_4\times S^7$ and $AdS_7\times S^4$ superspaces.

Notice that all the terms with background contributions to the {\it r.h.s.}'s of our mM$0$ equations are linear in fluxes. This is in disagreement with expectations based on the study of the Myers-type actions \cite{Myers:1999ps,YLozano+=0207}. Although the Myers action is purely bosonic and resisted all the attempts of its straightforward supersymmetric and Lorentz covariant generalization eleven years (except for the cases of lower dimensional and lower co-dimensional Dp-branes \cite{Dima01}) a particular progress in this direction was reached recently,  in the frame of the boundary fermion approach \cite{Howe+Linstrom+Linus} which gives a supersymmetric and Lorentz covariant description of multiple Dp-brane system but on a 'minus one-quantization' level (the quantization of boundary fermion sector is needed to be done to arrive at `classical' action). Taking this and also evidences from the string amplitude calculations into account, we do not exclude the possibility that the above mentioned discrepancy implies that our approach gives only an approximate description of the Matrix model interaction with supergravity fluxes (but, if so, it is Lorentz covariant, supersymmetric and going beyond the $U(N)$ SYM approximation).

If this is the case, a way to search for a more general interaction lays through modification of the basic equations of the superembedding approach of \cite{mM0=PLB,mM0=PRL}, namely the superembedding equation, defining the embedding of the mM$0$ center of energy superspace ${\cal W}^{(1|16)}$ into the target $D=11$ supergravity superspace $\Sigma^{(11|32)}$, and  the basic constraints of the $d=1$, ${\cal N}=16$ SYM model on the center of energy superspace. (Notice that the modification of the basic superembedding-type equations in the boundary fermion approach were suggested recently  in \cite{Howe+Linstrom+Linus=2010}). The problem of the deformation of the basic constraint determining the equations for the relative motion of mM$0$ constituents is the curved superspace generalization of the studies in \cite{Cederwall:2001td,Movshev:2009ba,Howe+Linstrom+Linus=2010}. However, unfortunately, if we allow consistent deformations of the basic equations of our superembedding approach, although  most probably these exist, they would certainly make the equations very complicated up to being unpractical.

Interestingly enough, if we do not deform the superembedding equation, but allow for a deformation of the
$d=1 {\cal N}=16$ $SU(N)$ SYM constraints on the center of mass superspace (see \cite{mM0=PLB,mM0=PRL}), the situation seems to be much more under control due to the rigid structure of the mM$0$ equations \cite{mM0=PRL}. In this case the center of mass motion and supersymmetry of the corresponding superspace is influenced only by the projections (\ref{M0:fFluxe}) of the supergravity fluxes, so that it is reasonable to assume that only these fluxes can enter the equations of  relative motion of the mM$0$ constituents. Then, as we have already noticed, the requirement of SO(1,1)$\times$SO(9) symmetry leaves very few possibilities to add the new terms to the ones already present in the {\it r.h.s.}'s of the equations (\ref{M0:DtPsi=})--(\ref{M0:DDXi=}). In particular, the only possible nonlinear contribution to the {\it r.h.s.} of the bosonic equations (\ref{M0:DDXi=}) is ${\bf X}^j \hat{F}_{\# jkl} \hat{F}_{\# ikl}$ describing the contribution  proportional to the second power of the 4-form flux to the mass matrix of the $su(N)$-valued fields ${\bf X}^j $.

%{\bf Acknowledgements.}
\ack{The author is grateful to Y. Lozano for useful discussions/communications. This work  was supported in part by the research grants FIS2008-1980 from the MICINN of Spain and the Basque Government Research Group Grant ITT559-10.}

\end{document}